\documentclass{article}
\usepackage{iclr2027_conference,times}
\usepackage[T1]{fontenc}
\usepackage{graphicx,booktabs,amsmath,amssymb,xcolor,microtype,wrapfig}
\usepackage{tabularx,xltabular,ragged2e}
\usepackage{algorithm,algpseudocode}
\usepackage{placeins}
\usepackage[hidelinks]{hyperref}
\usepackage{xurl}
\providecommand{\doi}[1]{doi:\ \href{https://doi.org/#1}{\begingroup\urlstyle{rm}\nolinkurl{#1}\endgroup}}
\newcolumntype{L}{>{\RaggedRight\arraybackslash}X}
\graphicspath{{figures/}}

\title{Cooldown Landmines: Cross-Tenant\\Interference Attacks on LLM Gateways}
\iclrfinalcopy
\author{\normalfont
\begin{minipage}{\dimexpr\textwidth-2\tabcolsep\relax}\centering
Yudong Gao$^{1,*}$, Linghan Chen$^{2,*}$, Wenhan Wu$^{3}$, Quan Shi$^{4}$, Xutao Mao$^{5}$, Mia Zhou$^{6}$,\\
Junjian Li$^{7}$, Xiaolong Liu$^{8}$, Jiyao Wang$^{9}$, Mingyu Guo$^{2}$, Honglong Chen$^{10,\dagger}$\\[5pt]
\normalsize
$^{1}$The Hong Kong University of Science and Technology \quad $^{2}$University of Adelaide\\
$^{3}$Wuhan University \quad $^{4}$National University of Singapore \quad $^{5}$City University of Hong Kong\\
$^{6}$University of North Carolina at Chapel Hill \quad $^{7}$Geely \quad $^{8}$Hunan University\\
$^{9}$ETH Zurich \quad $^{10}$China University of Petroleum\\[4pt]
\footnotesize $^{*}$Equal contribution.\quad $^{\dagger}$Corresponding author.
\end{minipage}%
}
\date{}
\hypersetup{pdftitle={Cooldown Landmines: Cross-Tenant Interference Attacks on LLM Gateways},pdfauthor={Yudong Gao, Linghan Chen, Wenhan Wu, Quan Shi, Xutao Mao, Mia Zhou, Junjian Li, Xiaolong Liu, Jiyao Wang, Mingyu Guo, Honglong Chen}}

\begin{document}
\maketitle
\lhead{}\chead{}\rhead{}\renewcommand{\headrulewidth}{0pt}

\begin{abstract}
LLM gateways enforce separate tenant quotas while sharing model deployments and cooldown records that temporarily exclude failing backends. However, a tenant's request failure can update these shared records and restrict other tenants' access to serviceable deployments. We identify two attacks that exploit this gap in LiteLLM. The first uses requests rejected at the key's requests-per-minute (RPM) limit: caller-supplied identifiers for known registered deployments reach failure handling, allowing two rejected requests to redirect another tenant to fallback with zero upstream calls from those requests. The second uses admitted traffic to create cooldown records that persist after backend capacity recovers. To address these failures, we design an origin check that blocks deployment updates from key RPM rejections and tenant-scoped cooldown that preserves the triggering tenant's back-off while retaining shared records for backend faults. Experiments with authenticated proxies and self-hosted vLLM demonstrate that the attacks can force fallback or denial while deployments remain serviceable. Across five paired four-worker runs, tenant scoping reduces victim fallback after recovery from $54/60$ to $0/60$, while increasing attempts against exhausted shared quotas. These findings show that tenant isolation must cover both request admission and the failure handling that governs shared deployment availability.
\end{abstract}

\section{Introduction}
An LLM gateway routes requests from a common API to model service endpoints, called deployments. Users or teams, called tenants, authenticate with separate keys and receive usage quotas while sharing provider or self-hosted deployments. This pooling lets tenants use capacity others leave idle. For example, the University of Minnesota uses LiteLLM with team keys, budget controls, and chargeback billing~\citep{umngateway}.

Sharing failure information can also improve reliability. If tenant $A$'s requests reveal a failed deployment, a cooldown record excludes it until expiry, letting $B$ use another deployment and avoid repeated failures. This benefit assumes the failure remains relevant to $B$. A tenant-specific limit or recovery before expiry can break that assumption, creating an isolation concern also found in adaptive clients~\citep{awsretry} and error caches~\citep{cpdos}.

In LiteLLM~\citep{litellm}, enforcing $A$'s key quota can itself generate a failure event that changes $B$'s routes. A request rejected at the key's requests per minute (RPM) limit retains a caller-supplied, known registered deployment ID in failure logging. The gateway then updates that deployment's shared cooldown without contacting it. A second path uses admitted requests to trigger HTTP~$429$ rate-limit errors at a deployment; the resulting cooldown can persist after capacity recovers. We study this cross-tenant interference by tracing which failures update deployment state and which tenants subsequently read it.

\begin{wrapfigure}{r}{.49\textwidth}
\centering
\includegraphics[width=\linewidth]{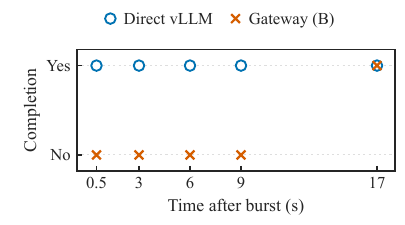}
\caption{\textbf{Exclusion after recovery.} Direct vLLM probes follow gateway calls and bypass the rate-limit layer. At $3$, $6$, and $9$\,s, the quota window is empty but $B$ is denied. One run, $15$\,s cooldown, fallback disabled (\S\ref{sec:causal}).}\label{fig:intro-recovery}
\end{wrapfigure}

Figure~\ref{fig:intro-recovery} illustrates the second path on vLLM~\citep{vllm}, an inference engine, with a layer enforcing request quotas before inference. At $3$, $6$, and $9$\,s after $A$ stops, the upstream's quota window (the interval for counting requests) is empty and direct vLLM calls return completions. Yet LiteLLM's unmodified (stock) routing logic excludes both primary deployments for $B$ because cooldown records persist. Tenant scoping serves $B$ at those samples in a separate run. We call this residual exclusion a \textbf{cooldown landmine}: $A$'s influence outlasts its traffic and the capacity restriction.

Fallback lets the gateway try an alternative group when the primary group is unavailable~\citep{litellmreliability}. Suppose $B$ permits requests only within a specified region. If cooldown excludes every eligible deployment there, successful fallback crosses that region boundary; enforcing the policy or disabling fallback leaves $B$ denied. Voting applications aggregate multiple model responses~\citep{agentforest}. Suppressing peer voters can let the remaining response determine a decision. We therefore measure denial, fallback routing, destination policy violations, and downstream voting behavior separately.

To test these mechanisms, we intervene separately on \emph{which failures may update deployment state} and \emph{which tenants read a cooldown record}. A server marker blocks deployment updates from key RPM rejections. Tenant scoping preserves $A$'s cooldown and retry delay (back-off) for eligible $429$ errors while letting $B$ try independently. Non-$429$ errors retain shared cooldown. These interventions measure the service benefit and the cost of independent attempts while shared capacity remains exhausted.

Our contributions are as follows:
\begin{itemize}
\item We identify shared cooldown state as an attack surface and demonstrate two attacks: locally rejected requests carrying registered deployment IDs, and admitted traffic that leaves deployments excluded after capacity recovers (\S\ref{sec:attack}).
\item We design two mitigations. The first prevents key RPM rejections from updating deployment failure state. The second scopes eligible $429$ cooldown records by authenticated tenant while retaining shared records for backend faults (\S\ref{sec:defense}).
\item We evaluate the attacks through authenticated proxy experiments and real vLLM probes, measuring fallback, denial, destination-policy violations, downstream voting effects, and mitigation costs. Across five paired four-worker runs, tenant scoping reduces victim fallback after recovery from $54/60$ to $0/60$ while retaining attacker back-off (\S\ref{sec:eval}).
\end{itemize}

\section{Related Work}\label{sec:related}
\textbf{Tenant isolation and failure handling.} Cloud systems isolate access to pooled resources~\citep{awstenantisolation} and allocate bandwidth to preserve tenant guarantees under contention~\citep{faircloud}. Failure handling also shares observations across requests. RFC~8084 requires protection of feedback authenticity and integrity~\citep{rfc8084}. AWS's adaptive limiter shares throttling within a client and bounds retries~\citep{awsretry}. LLM gateways combine tenant admission, deployment routing, and failure feedback. Our mitigations separate a failure's origin from the tenants that inherit its cooldown.

\noindent\textbf{Routing attacks and cached errors.} Learned routers trade response quality against inference cost~\citep{routellm}. Crafted inputs can manipulate their model selection~\citep{rerouting,lifecycle,routetorome}, motivating RerouteGuard~\citep{rerouteguard}. CPDoS~\citep{cpdos} stores attacker-induced error responses that later users receive. PROMPTPEEK~\citep{promptpeek} and KVGov~\citep{kvcache} study confidentiality and isolation in inference caches. We study deployment eligibility: a locally rejected request or an admitted deployment error updates cooldown state that a later tenant's router reads.

\noindent\textbf{Persistence and service consequences.} ContinuityBench~\citep{continuitybench} reports races in shared failover history and retries that sustain rate limiting. Shrew~\citep{shrew} exploits TCP retransmission timing, while metastable failures~\citep{metastable-hotos,metastable-osdi} use feedback that sustains failure after a trigger. A cooldown record has a finite lifetime but can outlive the load that created it, and later bursts can renew it (Appendix~\ref{sec:amplification}). Deployment constraints studied by ComplianceGate~\citep{compliancegate} show why the resulting route matters: fallback can cross a victim's residency or model boundary.

\section{System and Threat Model}\label{sec:threat}
\subsection{Gateway, deployments, and shared state}
An LLM gateway selects eligible deployments from a model group and may continue through configured fallback groups. Stock LiteLLM v1.99.0 stores one cooldown record per deployment, shared by tenants using the same cache. A qualifying failure updates this record, and later routing reads it to exclude the deployment. The default interval is $5$\,s. A $429$ can trigger immediate cooldown when the group has alternatives; explicit thresholds and single-deployment groups change this behavior. Permissions, region checks, and health checks further restrict the eligible set. Workers can share cooldown records through Redis.

\subsection{Adversary and isolation goal}
Attacker $A$ holds an ordinary valid tenant key and controls the timing and contents of its requests. It cannot access victim credentials, change gateway configuration, or write directly to the cooldown cache. Victim $B$ is another authenticated tenant that shares at least one deployment and its cooldown state with $A$; $B$'s requests pass its own authentication and quota checks. Experiments with admitted traffic enforce or explicitly bound $A$'s provisioned budget. Prompt content remains benign.

We study whether a failure caused by $A$ changes the deployments available to $B$. Such a change is appropriate when the failure identifies a condition that also applies to $B$, such as a backend outage. Cross-tenant interference arises when a tenant-specific failure, or a stale record after recovery, removes a deployment that could serve $B$. The local-rejection path assumes that $A$ knows registered target deployment IDs, which our harness supplies. A response header can expose a selected deployment ID; target discovery and its preparation cost remain unmeasured.

\section{Attacks on Shared Cooldown State}\label{sec:attack}
Our attacks aim to exclude serviceable deployments from victim $B$'s routes. Attacker $A$ induces failures that update shared cooldown records, which subsequently restrict $B$'s routing choices (Figure~\ref{fig:method-isolation}). We examine two paths. First, requests rejected at $A$'s key RPM limit use known registered deployment IDs to create records without contacting those deployments (\S\ref{sec:local-rejection}). Second, admitted traffic triggers deployment rate limits, leaving records that can persist after capacity recovers (\S\ref{sec:vuln}). Both paths can cause fallback, policy violations, or denial, depending on $B$'s remaining routes and fallback policy (\S\ref{sec:policy-consequence}).

\subsection{Local rejection with a caller-named deployment}\label{sec:local-rejection}
Admission rejection stops provider execution but leaves request metadata available to failure reporting. $A$ therefore exhausts its key allowance with admitted requests to a separate model, then uses the same key to send a request whose metadata names a known deployment used by $B$.

The proxy rejects this request with HTTP~$429$ but passes the supplied ID to deployment failure handling. Let $D$ be the deployments, $C(d)$ the shared cooldown record for $d\in D$, and $e$ the resulting failure event. Write $\widehat d(e)\in D$ for the supplied ID, $N(e)$ for the rejected request's upstream call count, and $r_e$ for its new cooldown record. Superscript $+$ denotes state after processing $e$. When the configured cooldown criteria are met, stock handling produces
\begin{equation}\label{eq:rejection-write}
 N(e)=0,\qquad C^+\!\left(\widehat d(e)\right)=r_e.
\end{equation}
Thus $B$'s next routing lookup can exclude an endpoint untouched by the rejected request. In our two-primary setting, one rejected request per registered primary redirects $B$ to fallback; missing and unknown IDs leave its route unchanged (\S\ref{sec:origin-attack}). The earlier admitted warmup reaches a provider.

\subsection{Admitted traffic and residual exclusion}\label{sec:vuln}
Tenant and deployment limits are separate, so $A$ can stay within its key allowance while exhausting a provider quota or deployment RPM limit. Qualifying $429$s create cooldown records and send later requests toward other candidates. Exclusion can cover the pool when $A$'s budget, routing, and quota windows allow these records to overlap (Appendix~\ref{app:quota-timing}).

The attack's persistence follows from two independently determined times: quota recovery and cooldown expiry. After the triggering limit clears, its record may still exclude a serviceable deployment. For one episode at $d$, let $\tau_{\mathrm{clear}}(d)$ be the limit's clearance time and $\tau_{\mathrm{expire}}(d)$ the record's expiry time, in seconds on the same clock. If clearance precedes expiry, residual exclusion can occur during
\begin{equation}\label{eq:residual-interval}
 I_d=[\tau_{\mathrm{clear}}(d),\tau_{\mathrm{expire}}(d)).
\end{equation}
At time $\tau\in I_d$, an otherwise eligible deployment that can serve $B$ remains excluded. We check serviceability, quota state, and cooldown at matched sample times (\S\ref{sec:causal}). Direct probes instrument our infrastructure; $A$ interacts through the gateway.

\begin{figure}[!htbp]
\centering
\includegraphics[width=\textwidth]{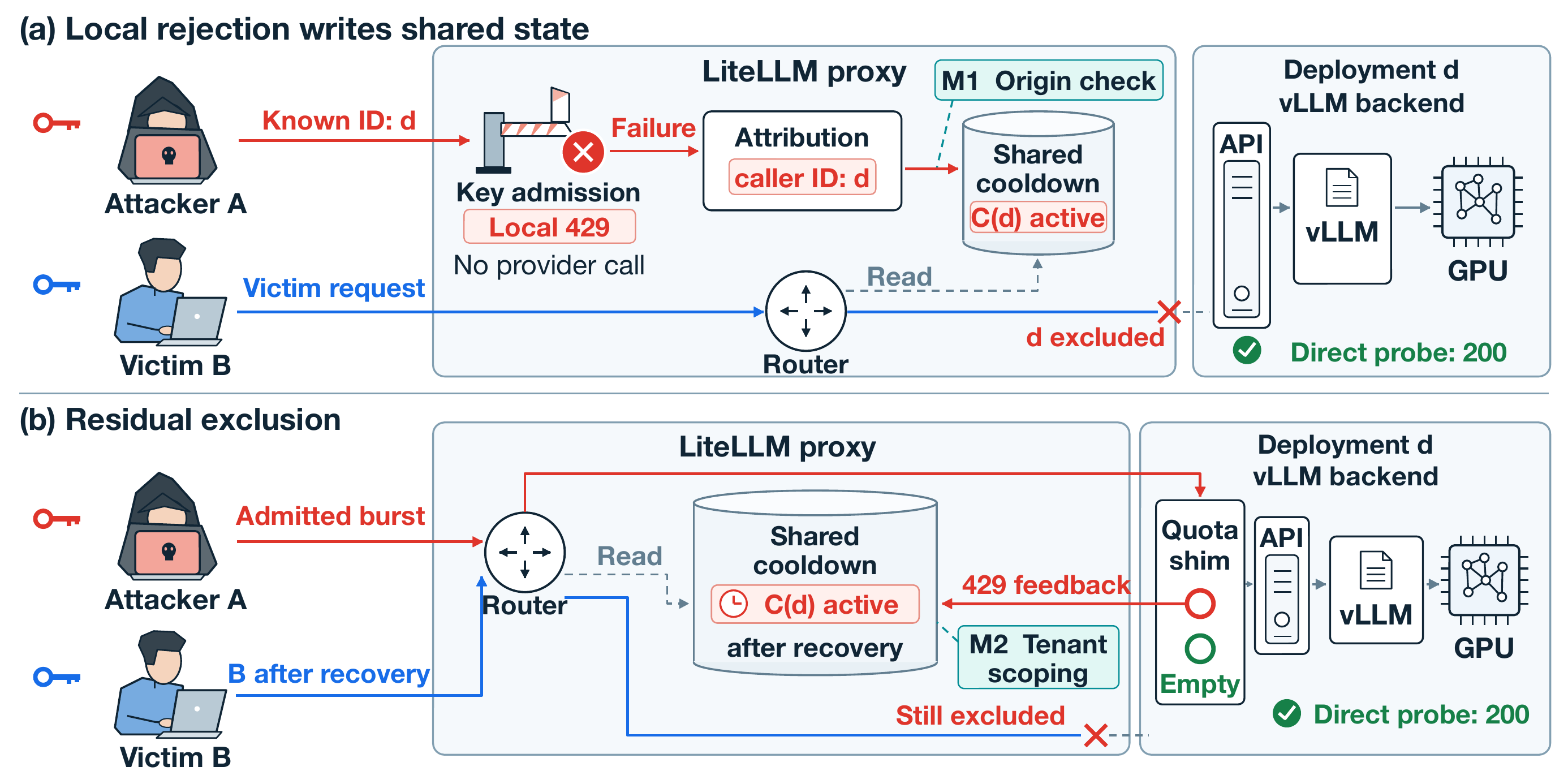}
\caption{\textbf{Attack paths and mitigation points.} Red paths originate from attacker $A$; blue paths show victim $B$; dashed gray arrows are cooldown lookups. (a) A key RPM rejection retains caller-named deployment $d$, updates shared record $C(d)$ without a provider call, and excludes the serviceable deployment from $B$. M1 marks the failure-origin check. (b) Admitted traffic receives a deployment $429$ and leaves $C(d)$ active after the quota window recovers, so $B$ remains excluded while a direct probe succeeds. M2 marks tenant-scoped storage for eligible $429$ records. The interventions are evaluated separately.}\label{fig:method-isolation}
\end{figure}

\subsection{From route exclusion to victim impact}\label{sec:policy-consequence}
At lookup time $\tau$, let $H_B(\tau)\subseteq D$ contain the serviceable deployments reachable through $B$'s primary and fallback groups that satisfy its destination requirements and all non-cooldown filters. Let $K(\tau)\subseteq D$ contain deployments with active shared cooldown. Removing them leaves the route set
\begin{equation}\label{eq:remaining-routes}
 R_B(\tau)=H_B(\tau)\setminus K(\tau),\qquad
 R_B(\tau)=\varnothing\ \Longleftrightarrow\ H_B(\tau)\subseteq K(\tau).
\end{equation}
For nonempty $H_B(\tau)$, this condition means cooldown removes every serviceable compliant route. Enforcing $B$'s requirements then yields denial. Permitting fallback outside those requirements can produce a residency or model-policy violation; a surviving route in $R_B(\tau)$ permits compliant routing.

\section{Mitigations}\label{sec:defense}
Section~\ref{sec:attack} shows that key RPM rejections can create deployment records, while records from admitted traffic can exclude other tenants after capacity recovers. We therefore control which failures update deployment state and which tenants read it (Figure~\ref{fig:method-isolation}). M1 blocks deployment updates from key RPM rejections. M2 scopes eligible $429$ records to the authenticated tenant, preserving its back-off while allowing other tenants to retry independently. Backend faults retain shared cooldown.

\subsection{Blocking deployment updates from key RPM rejections}\label{sec:origin}
Key RPM rejections and deployment errors reach LiteLLM's failure handlers as HTTP~$429$, so status alone does not establish the failure's origin. The gateway therefore marks the original key RPM exception before preparing failure metadata or invoking these handlers. For event $e$, let $p(e)=1$ when its exception carries this server-created marker, and $p(e)=0$ otherwise. Handlers check object identity within the gateway process, so client-supplied JSON cannot forge the marker.

Let $S$ comprise the deployment failure counters, deployment RPM accounting, and cooldown records touched by these handlers. If $\mathcal{U}_0(S,e)$ is their original state update, the guarded update $\mathcal{U}(S,e)$ is
\begin{equation}\label{eq:origin-update}
 \mathcal{U}(S,e)=
 \begin{cases}
 S, & p(e)=1,\\
 \mathcal{U}_0(S,e), & p(e)=0.
 \end{cases}
\end{equation}
This blocks the write in Eq.~\ref{eq:rejection-write} and prevents rejected requests from contributing to later failure thresholds. Tenant quota accounting and rejection logs remain outside $S$. Unmarked deployment RPM rejections and provider errors follow $\mathcal{U}_0$; a locally enforced deployment limit concerns that deployment even without a provider call.

\subsection{Scoping HTTP 429 cooldown by authenticated tenant}
For admitted traffic, a deployment $429$ warrants back-off by the triggering tenant, yet can outlive the capacity restriction that justified it. Shortening cooldown releases both $A$ and $B$ sooner. To let $B$ try independently while preserving $A$'s back-off, we instead separate tenant rate-limit records from shared fault records. For authenticated tenant $t$ and deployment $d$, let $G(d)$ be the global record and $Q(d,t)$ the tenant record; either may be absent. For an event $e$ meeting the cooldown criteria, let $s(e)$ be its integer HTTP status. The new record $r_e$ is written as
\begin{equation}\label{eq:scoped-write}
 \bigl(G^+(d),Q^+(d,t)\bigr)=
 \begin{cases}
 \bigl(G(d),r_e\bigr), & s(e)=429,\\
 \bigl(r_e,Q(d,t)\bigr), & s(e)\ne429.
 \end{cases}
\end{equation}
All other records remain unchanged by this write. Hence a qualifying $429$ updates the triggering tenant's record, while other errors retain global cooldown.

Routing then reads the tenant and global records together. For record $r$, let $r.\tau$ be its write time and $r.\Delta$ its duration in seconds. Write time and lookup time $\tau$ use the same clock. The predicate $\operatorname{live}(r,\tau)$ is $1$ if $r$ exists and $\tau<r.\tau+r.\Delta$, and $0$ otherwise. The exclusion indicator $C_t(d,\tau)$ is
\begin{equation}\label{eq:scoped-view}
 C_t(d,\tau)=\operatorname{live}(G(d),\tau)\lor\operatorname{live}(Q(d,t),\tau).
\end{equation}
Here $\lor$ denotes logical OR: either live record excludes $d$. For a qualifying $429$ from $A$, Eq.~\ref{eq:scoped-write} preserves $G(d)$ and $Q(d,B)$ whenever $A\ne B$. Comparing the states at the same lookup time, with other events fixed, therefore gives
\begin{equation}\label{eq:tenant-isolation}
 C_B^+(d,\tau)=C_B(d,\tau),\qquad A\ne B.
\end{equation}
Scoped routing replaces $K(\tau)$ in Eq.~\ref{eq:remaining-routes} with $K_B(\tau)=\{d\in D:C_B(d,\tau)=1\}$. For fixed $H_B(\tau)$, $A$'s write therefore preserves $B$'s remaining routes while retaining $A$'s back-off. A live global record still excludes the deployment for all tenants.

HTTP~$429$ alone does not identify the quota's owner~\citep{rfc6585}. Our policy scopes all qualifying $429$s by tenant, allowing additional attempts while a shared quota remains exhausted. Coordinated back-off would require a trusted mapping from tenants to provider quota owners. Section~\ref{sec:baseline} measures this cost and compares scoping with shorter cooldown.

\subsection{Identity and cache semantics}
Scoping requires failure writes and later routing reads to identify the same tenant, even across workers. Each operation therefore reads the registered key name, or otherwise the user ID, from authenticated request context. Distinct tenants need distinct identifiers, and each tenant needs a consistent identifier across its keys and gateway workers. Requests missing both fields share a default tenant namespace and therefore share its cooldown. This namespace remains separate from global error records. Appendix~\ref{sec:identity-eval} evaluates attribution across workers and concurrent tasks.

Tenant and global records use disjoint cache keys. The implementation preserves failure thresholds and supplied cooldown durations. A lookup batches one tenant and one global read per candidate and uses the later live expiry for exclusion and retry timing. Thus a shorter record cannot release a deployment while the other still excludes it. Appendix~\ref{app:cache-implementation} gives the algorithm and costs; \S\ref{sec:operational-main} measures latency, upstream requests, and Redis memory.

\section{Evaluation}\label{sec:eval}
We first test attack reachability, persistence, and impact, then mitigation effectiveness. We next measure operational costs and examine other configurations and gateways.

\subsection{Experimental setup}\label{sec:eval-setup}
We evaluate LiteLLM v1.99.0's Python \texttt{Router} and HTTP proxy using OpenAI-API mocks with controllable quotas and errors. Real inference uses Qwen1.5-MoE-A2.7B-Chat~\citep{qwenmoe} on vLLM 0.28.0, with a recoverable rate-limit layer for causal probes and two NVIDIA H800 endpoints for deployment-RPM experiments. The downstream voting case study uses DeepSeek-MoE-16B-Chat~\citep{deepseekmoe} through the same LiteLLM/vLLM path. US/EU denote local endpoint policies. Authenticated proxy tests store keys in PostgreSQL; four-worker runs share cooldown state through Redis. Appendix~\ref{app:extended-evaluation} gives configurations, protocols, and complete measurements.

\subsection{Attack evaluation}\label{sec:attack-eval}\label{sec:origin-attack}\label{sec:quota}\label{sec:attack-results}\label{sec:causal}\label{sec:downstream-voting}
\noindent\textbf{Local rejection.} We test whether a key RPM rejection can change another tenant's route. The proxy uses one worker, usage-based routing, $20$\,s cooldown, and no retries. One admitted request to a separate model exhausts $A$'s $1$\,rpm allowance. Two requests with absent or unknown deployment IDs and two naming registered primaries all receive local $429$s, with zero upstream calls. The first pair leaves $B$ on its primary; the registered IDs redirect $B$ to fallback (Table~\ref{tab:origin-controls}). Thus, a request rejected before provider execution can still select the deployment record that changes another tenant's route.

\noindent\textbf{Admitted traffic.} We next test whether $A$ can exhaust shared capacity within its key allowance. A routing-library test provisions $24$ requests per tenant against pool capacity $16$/window: $A$ stays within its allowance and denies idle $B$. In the authenticated proxy, each key allows $15$\,rpm while two primary deployments allow $6$\,rpm each. All $14$ requests from $A$ pass admission ($9$ US, $5$ EU), yet $B$ moves from US to EU after $A$ exceeds the primary pool's nominal $12$\,rpm capacity (Appendix~\ref{sec:reach-results}, Table~\ref{tab:twokey}). Therefore, admitted traffic can still trigger shared cooldown.

\begin{figure}[!htbp]\centering
\includegraphics[width=\linewidth]{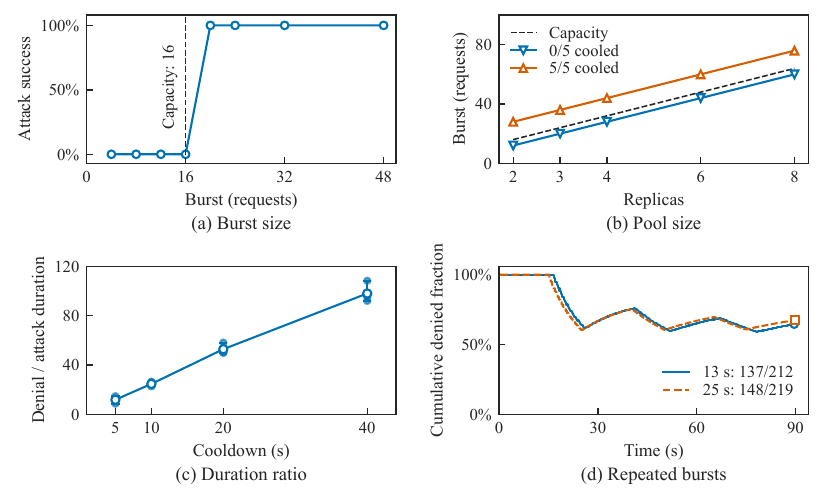}
\caption{\textbf{Attack reachability and persistence.} Admitted traffic in tests of the routing library with localhost mocks and no enforced key quotas. (a) Rerouting success over ten trials per burst; dashed line marks capacity. (b) Bursts below and above capacity ($8$ requests per replica), five trials per setting. Lines guide the eye. (c) Denial/attack duration over four trials: dots show observations, open markers the mean, and bars min--max. (d) Cumulative denied probes, one $90$\,s run per schedule; probing pauses during bursts.}\label{fig:attack-results}\label{fig:reach}\label{fig:pool}
\end{figure}

\noindent\textbf{Capacity threshold.} We vary burst size and replica count to locate the triggering threshold. With two mocks each admitting eight requests per two-second window, bursts up to $16$ reroute in $0/10$ trials, whereas bursts from $20$ reroute in $10/10$. Across $2$--$8$ replicas, bursts four below aggregate capacity cool the entire pool in $0/5$ trials; bursts twelve above it do so in $5/5$ (Fig.~\ref{fig:attack-results}a,b). The required burst therefore grows with aggregate capacity.

\noindent\textbf{Persistence.} We measure how cooldown duration and repeated bursts extend victim denial. Mean denial/attack duration grows from $12\times$ at $5$\,s cooldown to $98\times$ at $40$\,s, over four trials per setting (Fig.~\ref{fig:attack-results}c); these ratios depend on localhost request latency. Over $90$\,s, bursts at $13$\,s and $25$\,s intervals yield $137/212$ and $148/219$ denied probes at approximately $4\%$ and $1.5\%$ attacker duty cycle (Fig.~\ref{fig:attack-results}d). Probes run between bursts, with no enforced key quota (Appendix~\ref{sec:amplification}). Thus, cooldown outlasts each burst and can sustain denial between repeated bursts.

\noindent\textbf{Exclusion after recovery.} We isolate cooldown exclusion from ongoing resource contention. Stock cooldown, disabled cooldown, tenant scoping, and skipping $429$ cooldown use the same mock configuration and attacker schedule. At $3$, $6$, and $9$\,s after the burst, direct probes return $200$ and the quota window is empty. Stock's $15$\,s cooldown still denies $B$ with a router rate-limit error, while all three interventions serve $B$ (Appendix~\ref{app:causal-controls}). Therefore, shared cooldown withholds access after upstream quota recovery.

\noindent\textbf{Real inference and routing outcomes.} We check exclusion with real inference and compare fallback outcomes. Direct vLLM completions succeed immediately after refused gateway requests, while stock excludes $B$ and tenant scoping serves it (Fig.~\ref{fig:intro-recovery}). Separate mock trials record direct $200$s alongside gateway errors in $20/20$ trials (95\% Wilson CI $[84,100]\%$). When all compliant routes are cooled, a noncompliant fallback yields $20/20$ destination-policy violations; disabling fallback yields $20/20$ denials. Two H800 endpoints also exhibit rerouting and denial (Appendix~\ref{sec:real}). Thus, cooling all compliant routes leaves denial or fallback outside the victim's policy.

\noindent\textbf{Downstream voting.} We test decision changes while keeping $A$'s ballot fixed. In a five-voter workflow, $A$ always votes $A$; $B$--$E$ query DeepSeek-MoE-16B-Chat through LiteLLM/vLLM. After clean voting on nine non-$A$ multiple-choice probes, $A$ exhausts a $1$\,rpm key and sends locally rejected requests naming $B$--$E$'s registered deployments. These requests add zero upstream posts but leave all four peers denied by cooldown. All eight clean non-$A$ pluralities change to $A$; the five initially correct decisions all become wrong. A fixed $3$-of-$5$ quorum produces $9/9$ abstentions (Appendix~\ref{app:downstream-voting}). Thus, $A$ controls the observed plurality by suppressing peers, while a fixed quorum yields abstention.

\subsection{Defense evaluation}\label{sec:defense-evaluation}\label{sec:origin-eval}\label{sec:isolation-results}\label{sec:joint-eval}\label{sec:baseline}
\noindent\textbf{Origin check.} We test whether the origin check blocks key RPM updates while retaining deployment cooldown. Under the same rejection schedule, the server marker blocks the deployment update and preserves $A$'s rejection and $B$'s primary route (Table~\ref{tab:origin-controls}). Unmarked deployment-RPM and upstream-$429$ controls still produce fallback (Appendix~\ref{app:configuration-controls}). Thus, the check stops this attack path while preserving cooldown for the deployment errors in these controls.

\noindent\textbf{Tenant scoping across workers.} We test isolation across four workers in five paired runs using enforced key quotas, PostgreSQL, Redis, and recoverable mocks. $A$ sends $12$ requests under a $15$\,rpm key; a separate $1$\,rpm key tests admission rejection. After both primaries recover, each run sends $12$ victim requests across three or four workers. Stock sends $54/60$ to fallback; scoping sends all $60$ to recovered primaries (Table~\ref{tab:joint-outcomes}). Timestamps link attacker writes to other workers' reads in $14$ cases across seven victim requests (Appendices~\ref{app:worker-diagnostic} and~\ref{sec:identity-eval}). Scoping therefore keeps $A$'s cooldown outside $B$'s routing across workers.

\begin{table}[!htbp]\small
\begin{minipage}[t]{.48\linewidth}\centering
\caption{Victim routing after two locally rejected requests.}\label{tab:origin-controls}
\begin{tabularx}{\linewidth}{@{}Lcc@{}}\toprule
& \multicolumn{2}{c}{\textbf{Victim route}} \\
\cmidrule(l){2-3}
Deployment IDs & Stock & Origin check \\\midrule
Missing or unknown & Primary & Primary \\
Registered & Fallback & Primary \\\bottomrule
\end{tabularx}
\par\smallskip\raggedright
Upstream calls from rejected requests: $0$ in both policies.
\end{minipage}\hfill
\begin{minipage}[t]{.48\linewidth}\centering
\caption{Victim requests sent to fallback after recovery; $12$ requests per run.}\label{tab:joint-outcomes}
\begin{tabular*}{\linewidth}{@{\extracolsep{\fill}}lrr@{}}\toprule
Run & Stock & Tenant scoped \\\midrule
1 & $11$ & $0$ \\
2 & $10$ & $0$ \\
3 & $12$ & $0$ \\
4 & $12$ & $0$ \\
5 & $9$ & $0$ \\\midrule
Total & $54/60$ & $0/60$ \\\bottomrule
\end{tabular*}
\end{minipage}
\end{table}

\noindent\textbf{Comparison with shorter cooldown.} We test whether victim recovery can preserve $A$'s back-off. Five repetitions compare stock $12$\,s, short $1$\,s, and scoped $12$\,s cooldown. After $30$ attacker requests and a $2.5$\,s wait, both quota windows are empty in all $15$ runs. Short cooldown and scoping restore $B$ in $5/5$ runs; stock serves none. Scoping retains $A$'s back-off in every run, while short cooldown releases $A$ with $B$ (Table~\ref{tab:baseline}; Appendix~\ref{app:additional-cost}). Therefore, scoping restores $B$ while preserving $A$'s cooldown.

\begin{table}[!htbp]\centering\small
\caption{Recovery and cost across five repetitions. Recovery columns count runs; rate-limit columns give median upstream requests/served requests, and $500$ counts upstream requests. Each schedule issues $60$ requests ($120$ for $500$). Scoped four-tenant ranges are $16$--$18$ upstream requests and $8$--$10$ served.}\label{tab:baseline}
\setlength{\tabcolsep}{4pt}
\begin{tabular*}{\linewidth}{@{\extracolsep{\fill}}lcccccc@{}}\toprule
& \multicolumn{2}{c}{\textbf{After recovery}} & \multicolumn{4}{c}{\textbf{Upstream load}} \\
\cmidrule(lr){2-3}\cmidrule(l){4-7}
Policy & $B$ served & $A$ back-off & No retry & 2 retries & 4 tenants & $500$ \\\midrule
stock, $12$\,s & $0/5$ & $5/5$ & $10/8$ & $10/8$ & $10/8$ & $10$ \\
short, $1$\,s & $5/5$ & $0/5$ & $20/14$ & $20/14$ & $20/14$ & $20$ \\
scoped, $12$\,s & $5/5$ & $5/5$ & $10/8$ & $10/8$ & $17/9$ & $10$ \\\bottomrule
\end{tabular*}
\end{table}

\subsection{Operational costs}\label{sec:cost-results}\label{sec:load-eval}\label{sec:operational-main}
\begin{figure}[!htbp]\centering
\includegraphics[width=.93\linewidth]{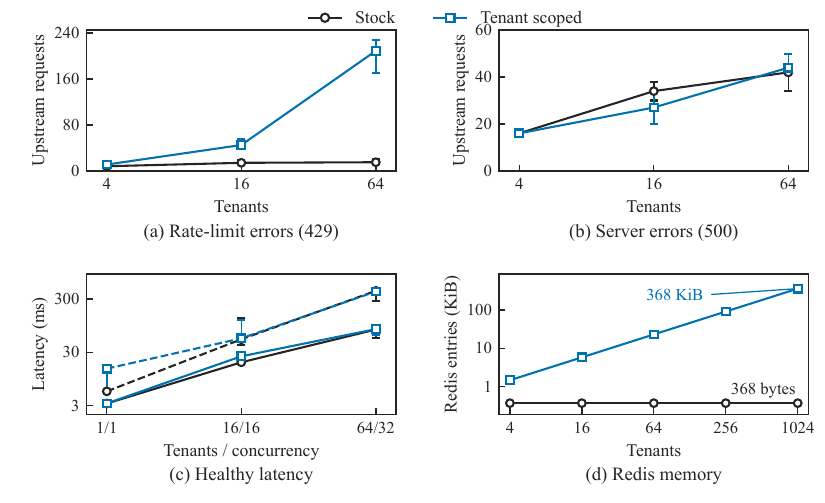}
\caption{\textbf{Cost of tenant scoping.} (a,b) Backend requests during four fault waves. (c) Healthy p50/p99 latency. Panels (a--c) show medians and observed min--max across five runs. (d) Redis memory for two deployments, one measurement per setting. Tenant counts use categorical positions.}\label{fig:operational-results}
\end{figure}

\noindent\textbf{Upstream requests.} We quantify the upstream cost of independent retries. In sequential schedules, short cooldown doubles attempts under overload and $500$ faults. Scoping matches stock's $10$ upstream attempts for one tenant, but reaches median $17$ with four tenants ($9$ served, $8$ rate-limited). Under concurrent $429$s at $64$ tenants, it raises the median from $15$ to $209$ requests while restoring $40/40$ probes after recovery, compared with $0/40$ for stock (Fig.~\ref{fig:operational-results}a). Thus, recovered access comes with more attempts against shared quotas.

\noindent\textbf{Backend faults.} We check global back-off and recovery for non-$429$ failures. Under $500$s at $64$ tenants, stock and scoping make median $42$ and $44$ upstream requests; both deny all $40$ immediate recovery probes (Fig.~\ref{fig:operational-results}b). At $4$ tenants, stock serves $35/40$ probes versus $13/40$ for scoping, which serves fewer in four paired runs and ties in one (Appendix~\ref{sec:concurrent-proxy}). Thus, both policies retain fault back-off, while the four-tenant result identifies an unresolved recovery difference.

\noindent\textbf{Latency and memory.} We measure healthy latency and retained cooldown memory. All $30{,}720$ healthy requests succeed. At $16$ tenants, median p50 rises from $19.35$ to $25.10$\,ms (Fig.~\ref{fig:operational-results}c); most requests use warm local caches and avoid Redis. With two deployments and $300$\,s TTLs, $1024$ tenants retain $2048$ scoped entries ($368$\,KiB), versus two stock entries ($368$ bytes) (Fig.~\ref{fig:operational-results}d). Therefore, scoping can add latency with warm caches and requires storage per tenant and deployment.

\subsection{Generalization}\label{sec:generalization}\label{sec:scope}
\noindent\textbf{Other configurations.} We inspect AWS's LiteLLM~1.89.1 example to check how configuration affects shared state (Table~\ref{tab:config-families}). RPM rejection before deployment selection bypasses cooldown; eligible failures attributed to a deployment can update it. A primary and fallback share one deployment ID despite different RPM settings; another fallback has a distinct key. Authentication is unspecified, so the sample supplies static key and rejection-path evidence (Appendix~\ref{sec:general}). Thus, a fallback with a different quota can inherit primary cooldown.

\noindent\textbf{Other gateways.} We test exclusion after recovery in Kong~3.6, where two authenticated users share backend health records. With passive health checking and a three-failure threshold, $A$ excludes backends that subsequently return direct $200$s while $B$ receives $503$. Passive checking is off by default; periodic backend probes (active checks) restore service after approximately $9$\,s in the observed run. Consequently, shared health records prolong exclusion until active probes restore service.

\subsection{Discussion}\label{sec:discussion}\label{sec:limitations}
We distinguish the properties established by the evaluation from those requiring production measurements. Sequential recovery tests isolate cooldown persistence; concurrent mock loads test shared state across workers. The origin check is evaluated for key RPM rejection in one worker. These results motivate production studies of sustained interference under enforced key quotas, exact RPM accounting across workers, provider quota ownership, dynamic credentials, patch propagation, and concurrent retries.

\FloatBarrier
\section{Conclusion}
LiteLLM's cooldown state creates cross-tenant interference when tenant failures update deployment records read by other tenants. Key RPM rejections can update caller-named deployments without provider calls. Admitted traffic can leave deployments excluded after capacity recovers. The consequences include fallback, denial, policy violation, and changed downstream decisions. A server marker blocks key-RPM updates. Tenant scoping reduces victim fallback from $54/60$ to $0/60$ across five paired runs while increasing attempts against exhausted shared quotas.

\clearpage
\subsection*{Ethics Statement}
All experiments ran on our own infrastructure against mock or self-hosted upstreams, including our two GPUs, using synthetic requests with benign content. The experiments involved no third-party or hosted targets and no real user data. We study the cross-tenant consequences of documented cooldown and health-check controls under specified configurations and evaluate a mitigation prototype. Kong's shared passive health checking is documented and depends on configuration, with active health checks bounding the observed exclusion window.

Maintainer disclosure is pending. We plan to disclose the cross-tenant impact and proposed mitigation before publication, allow a remediation window, and then release the test harness and prototype so operators can evaluate their own deployments.

\subsection*{AI Use Statement}
Generative AI tools assisted with refining the research framing and hypotheses, designing experiments and controls, implementing experimental harnesses and prototype code, analyzing logs and interpreting results, finding and summarizing related work, drafting and revising manuscript text, and creating and editing figures. The authors are responsible for the final text, claims, code, and figures.

\subsection*{Reproducibility Statement}
Section~\ref{sec:eval-setup} and Appendices~\ref{app:extended-evaluation} and~\ref{sec:general} specify software, authentication, quotas, backends, execution settings, and protocols alongside the results. To support reproducibility, we provide an anonymized repository containing the
implementation, pinned environments, experiment scripts, archived results, and
instructions for reproducing the evaluation:
\url{https://anonymous.4open.science/r/llm-gateway-cooldown-security-CE70/README.md}.

\nocite{yu2025blackbox,chen2026fingerprints,gao2026pathmark,li2026nddl,mao2026trustfork,mao2026cave}
\bibliographystyle{iclr2027_conference}
\bibliography{refs}

\clearpage
\appendix
\section{Cooldown Cache Implementation}\label{app:cache-implementation}
\subsection{Cooldown writes and routing reads}
We extend LiteLLM's cooldown cache, the component that stores deployment exclusion records, by subclassing \texttt{CooldownCache}~\citep{litellm}. The router's failure thresholds and trigger logic determine when to request a cooldown write. For each write, the subclass converts the status to an integer: $429$ selects the current tenant's records, and every other status follows the global path. Errors with $5xx$ status retain the shared cooldown path.

The authentication hook selects the registered key name (\texttt{key\_alias}), then \texttt{user\_id}, then the default tenant identifier \texttt{GLOBAL}. A Python \texttt{ContextVar} carries this identity through request execution for each cache operation to retrieve. Records under this default tenant identifier are separate from global error state.

Algorithm~\ref{alg:scoped-cache} summarizes the scoped cache using the deployment identifier $d$, tenant identifier $t$, time $\tau$, and record $r$ defined in \S\ref{sec:defense}. The cache $M$ maps keys to records. Both operations receive $t$ from the authenticated request context at entry. The procedure $\operatorname{RecordCooldown}$ takes $d$, $t$, error status $s$, and optional duration $\Delta$ in seconds. The function $\operatorname{status}(s)$ converts the status to an integer; $\Delta_{\mathrm{default}}$ supplies an absent duration. The functions $K_T(d,t)$ and $K_G(d)$ return keys for $Q(d,t)$ and $G(d)$; their subscripts label tenant and global records. The local variable $k$ holds the selected key, and $\operatorname{now}()$ returns a timestamp in seconds.

A write $M.\operatorname{put}(k,r,\mathrm{TTL}=\Delta)$ stores $r$ under $k$ for $\Delta$ seconds, replacing the previous value. The record fields $\tau$, $\Delta$, and $\mathrm{status}$ hold its write timestamp, duration, and error status. A read $M.\operatorname{get}(k)$ returns the record or an absent value. The local variables $r_t$ and $r_g$ hold tenant and global records; the subscript $g$ labels global state. The set $V$ contains their live records, and $\bot$ denotes absence. The operator $\arg\max$ selects a record with the latest expiry, preferring the global record on a tie. The assignment arrow $\gets$ stores its right-hand value in its left-hand variable.

\begin{algorithm}[htbp]
\caption{Cooldown writes, exclusion, and metadata selection}\label{alg:scoped-cache}
\small
\begin{algorithmic}[1]
\Require Authenticated tenant identity $t$; cache $M$
\Procedure{RecordCooldown}{$d,t,s,\Delta$}
  \If{$\Delta$ is absent}
    \State $\Delta\gets\Delta_{\mathrm{default}}$
  \EndIf
  \If{$\operatorname{status}(s)=429$}
    \State $k\gets K_T(d,t)$
  \Else
    \State $k\gets K_G(d)$
  \EndIf
  \State $r\gets(\tau=\operatorname{now}(),\Delta=\Delta,\mathrm{status}=s)$
  \State $M.\operatorname{put}(k,r,\mathrm{TTL}=\Delta)$
\EndProcedure
\Function{ActiveRecord}{$d,t,\tau$}
  \State $r_t\gets M.\operatorname{get}(K_T(d,t))$
  \State $r_g\gets M.\operatorname{get}(K_G(d))$
  \State $V\gets\{r\in\{r_t,r_g\}:\operatorname{live}(r,\tau)=1\}$
  \If{$V=\emptyset$}
    \State \Return $\bot$
  \EndIf
  \State \Return $\arg\max_{r\in V}(r.\tau+r.\Delta)$
\EndFunction
\Function{Excluded}{$d,t,\tau$}
  \State \Return $\operatorname{ActiveRecord}(d,t,\tau)\ne\bot$
\EndFunction
\end{algorithmic}
\end{algorithm}

The function $\operatorname{Excluded}$ returns the indicator $C_t(d,\tau)$ in Eq.~\ref{eq:scoped-view}: true is $1$ and false is $0$. Selecting the latest expiry preserves exclusion until both applicable records expire. The API returns a copy of the selected record's metadata with timestamp $\tau$ and remaining duration $r.\tau+r.\Delta-\tau$, leaving cached records unchanged. Thus, a short tenant cooldown cannot conceal a longer global fault, and a short global cooldown cannot shorten the triggering tenant's back-off. Retry timing uses these same records and returns the minimum remaining exclusion across requested deployments, including zero if any has no live record. Other routing filters can still prevent service.

The recovery, paired proxy, and policy comparisons use tenant-first metadata selection and LiteLLM's global-only lookup for retry timing. The concurrent proxy experiments use the latest-expiry selection in Algorithm~\ref{alg:scoped-cache} for synchronous reads, asynchronous reads, and retry timing. Both implementations apply the exclusion predicate in Eq.~\ref{eq:scoped-view}. Tenant keys include the deployment and tenant IDs; global keys include only the deployment ID.

Skipping cooldown writes for $429$ removes the record for all tenants, allowing later requests to reach the upstream during its rate-limit interval. Tenant scoping retains the triggering tenant's record and back-off. Section~\ref{sec:causal} compares recovery, and Appendix~\ref{app:additional-cost} reports the upstream load from skipping $429$ cooldown. Section~\ref{sec:load-eval} compares stock, short, and tenant-scoped cooldown under matched schedules.

\subsection{Lookup and storage cost}
Let $m$ be the number of deployment candidates in a lookup. Reading one tenant and one global record per candidate addresses at most $2m$ records. The asynchronous implementation batches these reads, pairs the returned records by deployment, and applies the expiry predicate from Eq.~\ref{eq:scoped-view}.

For deployment set $D$, let $T_{\mathrm{active}}$ be the tenants retaining at least one cooldown record for a deployment in $D$. The cache stores at most $|D|$ global records and $|D||T_{\mathrm{active}}|$ tenant records, for a worst-case $O(|D|+|D||T_{\mathrm{active}}|)$ record footprint. Records expire after their TTL or may be removed earlier by cache eviction. Early removal weakens back-off and may increase later upstream attempts.

\subsection{Quota ownership and recovery timing}\label{app:quota-timing}
The admitted path studies the effect of sharing failure feedback. A deployment may use a shared provider quota or a gateway limit applied to that deployment. Alternatively, a provider may enforce separate quotas for downstream tenants. If $A$ exhausts its own provider quota while $B$ retains allowance, a deployment-wide cooldown can still exclude $B$. The relevant condition is that a failure attributable to $A$ creates a record consulted by $B$. Enforcing separate quotas at the provider changes which requests would succeed; assigning separate cooldown records changes which requests the gateway permits to try.

Residual exclusion requires recovery before the applicable cooldown expires. A long quota window alone does not determine this ordering: a fixed minute window may reset one second after a failure, leaving four seconds of exclusion under a five-second cooldown. Conversely, when capacity remains unavailable throughout cooldown, that burst produces no exclusion after recovery. Repeated bursts can create further records, subject to admission budgets and failure thresholds. Appendix~\ref{app:extended-evaluation} specifies the quota windows and cooldown durations with each protocol.

The local-rejection path depends on LiteLLM copying the caller's top-level \texttt{model\_info} into the failure context, where the router reads \texttt{litellm\_params.model\_info.id}. The response header \texttt{x-litellm-model-id} can disclose the selected deployment's ID. Removing it reduces one source of disclosure; previously known or otherwise exposed IDs remain usable. Deployment-ID discovery remains unmeasured.

The marker is a server-owned enum member attached to the original exception. Before updating deployment state, the Router failure callback, asynchronous deployment RPM accounting callback, and common cooldown writer each check the marker by object identity.

\subsection{Thresholds and configuration alternatives}\label{app:configuration-controls}
LiteLLM v1.99.0 controls both cooldown triggers and duration. A deployment can use separate failure thresholds for rate-limit errors, server errors, and timeouts. Bursts below the threshold avoid that cooldown path; larger bursts or accumulated failures can cross it. A deployment's cooldown duration override applies across error types, with zero disabling the interval. Section~\ref{sec:baseline} measures recovery and load with a one-second override. Both configuration changes preserve the reader set.

Other controls act earlier or later in the attack sequence. Admission limits that cover the relevant windows and request amplification can prevent the trigger, while dedicated deployment pools remove shared deployment state. Once exclusion occurs, destination checks and an unaffected compliant fallback can prevent the policy consequence. Tenant scoping acts on the state dependency itself: a write attributed to $A$ is excluded from $B$'s view while $A$'s back-off remains active. Section~\ref{sec:baseline} evaluates that property and its costs.

\noindent\textbf{Unmarked failure controls.} Two Router controls check that the marker leaves deployment failures on their existing paths. With deployment \texttt{rpm=1}, the first request uses primary and the second uses fallback. An upstream returning $429$ is called once before the caller is routed to fallback. Both errors are unmarked.

\section{Extended Evaluation}\label{app:extended-evaluation}
Experiments using the routing library bypass proxy authentication. Recovery, destination policy, and cooldown policy tests use sequential requests without per-key quotas, as does routing on the two H800 endpoints. Threaded and asynchronous identity tests use the routing library with mocks, supplied tenant identities, and no per-key quotas. Proxy tests use the authentication and admission settings specified with their protocols below.

\begin{table}[htbp]\centering\small
\caption{Evaluation organized by the question each result answers. Settings and protocols accompany the results below.}\label{tab:eval-overview}
\begin{tabularx}{\textwidth}{@{}>{\RaggedRight\arraybackslash}p{3.4cm}LL@{}}\toprule
Result group & Question & Evidence \\\midrule
Attack impact and causality (\S\ref{sec:attack-eval}) & How do admitted traffic and local rejection change victim routing? & Probes after recovery, cooldown interventions, locally rejected requests, destination policy, and real vLLM outcomes. \\
Reachability and persistence (\S\ref{sec:quota}; Appendix~\ref{sec:reach-results}) & Which budgets and pool sizes admit the attack, and how long does interference last? & Enforced key quotas, burst and replica sweeps, cooldown timing, and repeated bursts. \\
Isolation effectiveness (\S\ref{sec:defense-evaluation}) & Do the origin check and tenant scoping isolate failures while retaining back-off? & Origin veto, five paired proxy runs with four workers, cooldown reads across processes, and tenant attribution. \\
Operational costs (\S\ref{sec:operational-main}) & How does isolation change upstream work, latency, and retained state? & Recovery, overload, retry, fault, and multiple tenants under the same schedules, repeated five times; concurrent proxy latency and measured Redis allocation. \\\bottomrule
\end{tabularx}
\end{table}

\subsection{Recovery protocols and controls}\label{app:causal-controls}
\begin{figure}[t]\centering
\includegraphics[width=0.80\linewidth]{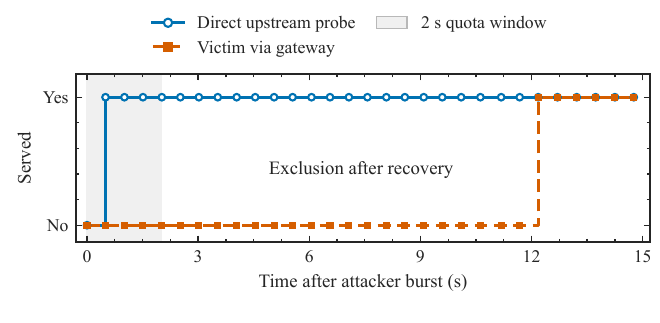}
\caption{After $A$'s burst, the upstream is healthy within ${\sim}0.5$\,s (direct probe $=200$). The gateway continues denying tenant $B$ for the full cooldown ($\approx$12\,s here), extending the harm beyond contention.}\label{fig:timeline}
\end{figure}

After the physical window recovers, a direct upstream probe returns $200$ while the gateway request for $B$ fails. This co-occurrence appears in $20/20$ trials (95\% Wilson CI $[84,100]\%$); the statistic counts gateway errors without distinguishing their exception classes. Figure~\ref{fig:timeline} records the separate timeline experiment, in which exclusion persists for approximately $12$\,s after a sub-second burst.

\noindent\textbf{Causal control (Table~\ref{tab:causal}).} Four mock arms share a base configuration and attacker schedule: stock cooldown, disabled cooldown, tenant scoping, and skipping $429$ cooldown. At nominal times $\tau=0.5,3,6,9,17$\,s after the burst ends ($\tau=0$), we record direct health, quota-window requests, the cooldown set, and the victim outcome with its full exception. Exclusion changes realized traffic by withholding requests. With cooldown disabled, $B$ encounters an upstream \texttt{RateLimitError} at $0.5$\,s and is served thereafter. Stock's $15$\,s cooldown denies at $3$, $6$, and $9$\,s despite upstream $200$ and zero occupancy, returning \texttt{Router\allowbreak Rate\allowbreak Limit\allowbreak Error}, then serves at $17$\,s. Scoped and skip-$429$ serve from $3$\,s. These probes identify exclusion after recovery at the sampled times.

\noindent\textbf{Control on real vLLM.} To check serviceability with real inference, we repeat the stock, disabled cooldown, and tenant scoping arms with a recoverable rate-limit layer before a live Qwen1.5-MoE vLLM server. Stock denies at samples after recovery while disabled cooldown returns a completion. Immediately after each gateway request, a paired probe bypasses rate limiting and cooldown to call vLLM directly. It returns real completions where stock denies $B$, establishing serviceability immediately after the recorded refusal. Figure~\ref{fig:intro-recovery} compares the stock gateway outcomes with these paired direct completions. At $3$, $6$, and $9$\,s, the quota window is empty and both primary deployments remain excluded; the separate scoped arm serves $B$ at all three samples.

\begin{table}[H]\centering\small
\caption{Causal control: victim with fallback disabled at $\tau{=}6$\,s after the burst ends, \emph{after} the physical window recovered (upstream $200$, occupancy $0$).}\label{tab:causal}
\begin{tabularx}{\columnwidth}{@{}LllL@{}}\toprule
arm & cooled set & victim & exception \\\midrule
stock ($15$\,s) & \{d1,d2\} & \textbf{denied} & \texttt{Router\allowbreak Rate\allowbreak Limit\allowbreak Error} \\
cooldown off (all) & \{\} & served & none \\
tenant-scoped ($429$) & per-tenant & served & none \\
$429\!\to$ no cooldown & \{\} & \textbf{served} & none \\\bottomrule
\end{tabularx}
\end{table}

\subsection{Availability and destination policy}\label{sec:tenant-controls}
We evaluate controls available to the victim in a pool where the attack cools every otherwise eligible compliant deployment and no unaffected compliant alternative is available ($20$ trials). Fallback to the noncompliant region yields a policy violation $20/20$, while disabling fallback yields denial $20/20$. A same-region fallback chain confined to the same cooled pool also leaves the victim denied. Table~\ref{tab:dilemma} shows how these controls select between the two outcomes.

\begin{table}[t]\centering\small
\caption{Victim controls when cooldown excludes all compliant deployments and no unaffected compliant alternative exists. Stock controls read the same cooldown record for each deployment.}\label{tab:dilemma}
\begin{tabularx}{\columnwidth}{@{}LLl@{}}\toprule
Victim $B$'s control & Destination policy & Service \\\midrule
Fallback enabled & Violated (cross-region) & Served \\
Fallback disabled & Preserved & Denied \\
Same-region fallback in cooled pool & Preserved & Denied \\\bottomrule
\end{tabularx}
\end{table}

\subsection{Downstream voting workflow}\label{app:downstream-voting}
The downstream case study uses the proxy with database-backed keys, DeepSeek-MoE-16B-Chat on vLLM, and two registered deployments per voter. Voters $B$--$E$ produce short answers parsed from an explicit final answer or the selected option text. Clean decisions use the successful-response plurality over five voters; the fixed-quorum variant requires at least three valid votes.

\begin{table}[t]\centering\small
\caption{Downstream voting after local-rejection cooldown updates. The attack suppresses voters $B$--$E$ through gateway cooldown records; $A$'s fixed ballot is unchanged.}\label{tab:downstream-vote}
\begin{tabularx}{.82\linewidth}{@{}Lr@{}}\toprule
Measure & Outcome \\\midrule
Target voters denied after forged local rejections & $4/4$ \\
Upstream posts from forged requests & $0$ \\
Clean non-$A$ decisions changed to $A$ & $8/8$ \\
Clean-correct non-$A$ decisions changed to $A$ & $5/5$ \\
Fixed $3$-of-$5$ quorum after attack & $9/9$ abstain \\\bottomrule
\end{tabularx}
\end{table}

\subsection{Outcomes on real vLLM}\label{sec:real}
Two independent vLLM 0.28.0 servers run Qwen1.5-MoE-A2.7B-Chat, each on its own NVIDIA~H800 (FLASH\_ATTN backend, CUDA graphs, 27.9\,GiB weights/GPU). LiteLLM's deployment \texttt{rpm} limiter triggers cooldown in front of these endpoints. Table~\ref{tab:vllm} records rerouting and denial with real inference. The causal control in \S\ref{sec:causal} measures serviceability after recovery using direct vLLM completions.

\begin{table}[t]\centering\small
\caption{Real inference on two local H800 endpoints running vLLM, with deployment RPM limits and no per-key quota. US/EU denote assigned policy regions.}\label{tab:vllm}
\begin{tabularx}{\columnwidth}{@{}LL@{}}\toprule
Stage & Victim $B$ outcome \\\midrule
baseline & served by endpoint US \\
after $A$'s burst, failover on & \textbf{rerouted to endpoint EU} \\
after $A$'s burst, fallback disabled & \textbf{denied} \\\bottomrule
\end{tabularx}
\par\smallskip{\footnotesize Denial in this test returns \texttt{ValueError} without a recorded cause chain; the causal control in Table~\ref{tab:causal} identifies \texttt{Router\allowbreak Rate\allowbreak Limit\allowbreak Error}.\par}
\end{table}

\subsection{Attack reachability and persistence}\label{sec:reach-results}
\begin{table}[htbp]\centering\small
\caption{Two virtual keys stored in the database, with mock backends and quotas enforced by the gateway: $15$ rpm per key, versus $6$ rpm per deployment and $12$ rpm across the primary pool. The attacker stays within its quota while the victim is rerouted.}\label{tab:twokey}
\begin{tabularx}{\columnwidth}{@{}LLL@{}}\toprule
 & key $A$ (attacker) & key $B$ (victim) \\\midrule
RPM per key (enforced by LiteLLM) & $15$/min & $15$/min \\
before attack & n/a & \texttt{US} \\
requests issued & $14$ (\textbf{all admitted}) & $1$ \\
served after $A$'s burst & $9$ \texttt{US} / $5$ \texttt{EU} & \textbf{\texttt{EU} (cross-tenant)} \\\bottomrule
\end{tabularx}
\end{table}
Table~\ref{tab:twokey} reports the attack under enforced tenant quotas. The following routing-library sweeps use mocks without per-key quotas to measure the burst needed to cover the pool and the duration of exclusion.

\subsubsection{Pool coverage and scaling}\label{sec:pool-scaling}
Figure~\ref{fig:attack-results}(a,b) shows how the required burst grows with pool capacity. The pool sweep tests $N$ replicas and capacity $C=8N$ requests per window; bursts $C-4$ and $C+12$ cool the entire pool in $0/5$ and $5/5$ trials, respectively.

\subsubsection{Denial duration and repeated attacks}\label{sec:amplification}
\begin{figure}[t]\centering
\includegraphics[width=\linewidth]{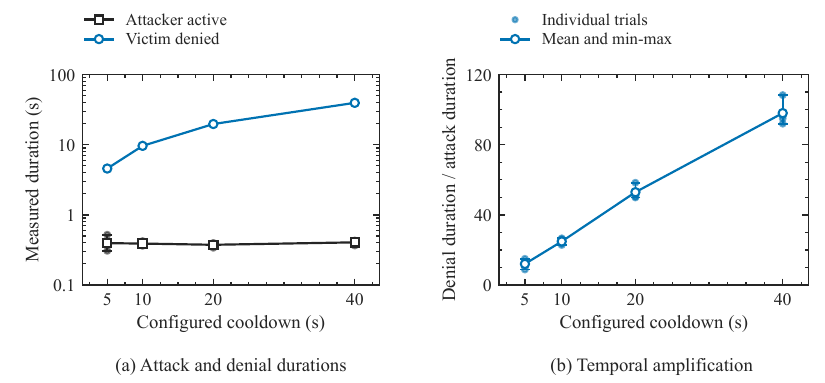}
\caption{Temporal amplification over $4$ trials per cooldown setting. (a) Attack and denial durations on a logarithmic axis. (b) The ratio of denial duration to attack duration in each trial. Small dots show individual trials; open markers and error bars show means and min--max ranges.}\label{fig:amp}
\end{figure}

Amplification is the ratio of measured denial duration to attack duration in the localhost experiments. It averages $\approx 12\times$ at $5$\,s cooldown and $\approx 98\times$ at $40$\,s over four trials per point (Fig.~\ref{fig:amp}); the latter range across trials is $92$--$108\times$. These timing ratios depend on localhost request latency.

\noindent\textbf{Trigger and duration.} Responses with status $429$ trigger cross-tenant cooldown in both the OpenAI and Anthropic API formats, including their error bodies and \texttt{Retry-After} headers specifying a suggested retry delay. Cooldown duration follows this precedence: per-deployment override, then provider \texttt{Retry-After}, then router default. With a $30$\,s header, an $8$\,s default, and no deployment override, victim denial lasts approximately $27$\,s. A deployment override therefore controls the recorded duration even when the provider supplies a longer value.

\begin{figure}[t]\centering
\includegraphics[width=\linewidth]{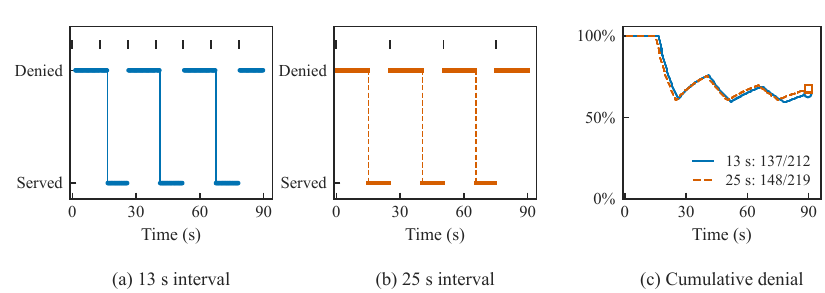}
\caption{Sustained cross-tenant denial over $90$\,s. (a,b) Outcomes of victim probes under $13$\,s and $25$\,s burst intervals, at approximately $4\%$ and $1.5\%$ attacker duty cycle. Black ticks mark attacker bursts. State traces connect adjacent observations and break when victim probing pauses. (c) Cumulative denied probes divided by all probes observed so far, ending at $137/212$ and $148/219$.}\label{fig:sustain}
\end{figure}

\noindent\textbf{Repeated triggering.} Over a $90$\,s observation window, periodic bursts yield $137/212$ denied victim probes under the aggressive schedule and $148/219$ under the sparse schedule, approximately $65$--$68\%$, at a $1.5$--$4\%$ attacker duty cycle (Fig.~\ref{fig:sustain}). The denominator counts victim requests sampled between attacker bursts. Renewal requires reaching a primary after recovery; requests sent to fallback affect the fallback path.

\subsection{Tenant isolation and fault back-off}\label{sec:defense-eval}
We compare skipping $429$ cooldown and tenant scoping with stock cooldown. Skipping $429$ cooldown serves the victim at each sample after recovery, starting at $3$\,s (Table~\ref{tab:causal}), removing residual cooldown for both the triggering tenant and other tenants. Tenant scoping preserves the triggering tenant's back-off while allowing a distinct victim to use recovered capacity. In the $429$ attack case, stock serves $B$ in $0/15$ trials versus $15/15$ for tenant-scoped cooldown. Under a genuine $500$ outage, both route $B$ away from the dead deployment (Table~\ref{tab:defense}).

\begin{table}[t]\centering\small
\caption{Mitigation 2$\times$2 in the evaluated cases: victim service after the rate-limit window recovers, and routing away from a backend returning $500$.}\label{tab:defense}
\begin{tabularx}{\columnwidth}{@{}LLL@{}}\toprule
 & 429 after $A$'s burst & 500 (genuine outage) \\\midrule
stock LiteLLM & \textbf{$B$ denied} & $B$ avoids dead $d$ \\
tenant-scoped fix & \textbf{$B$ served} & $B$ avoids dead $d$ \\\bottomrule
\end{tabularx}
\end{table}

\subsection{Worker traces and intervals between Redis reads}\label{app:worker-diagnostic}
\begin{table}[htbp]\centering\small
\caption{Log consistency across five paired runs with four workers. Entries show matched events / expected events. Per-run routing outcomes appear in Table~\ref{tab:joint-outcomes}.}\label{tab:joint-repeat}
\begin{tabularx}{\linewidth}{@{}Lrr@{}}\toprule
Check & Stock & Tenant scoped \\\midrule
HTTP requests recorded in access logs & $170/170$ & $170/170$ \\
LLM callbacks with process IDs matching the access log & $155/155$ & $155/155$ \\
Cooldown starts matched to request and authenticated tenant & $433/433$ & $219/219$ \\\bottomrule
\end{tabularx}
\end{table}
A four-worker stock proxy using Redis, mocks, a shared master key, and no per-key quota reroutes all $30$ victim probes after the attack. With database virtual keys and tenant admission limits (\S\ref{sec:joint-eval}), all $340$ logical requests appear in access logs, all $310$ LLM callback process IDs match those logs, and all $652$ cooldown start events match the request and authenticated tenant mapping (Table~\ref{tab:joint-repeat}).

In five stock runs with the same victim schedule, routing varies with when each worker can reread Redis. All $41/60$ fallback requests observe both US cooldowns on the first routing read. Each of the $19/60$ primary responses follows a read of exactly one; the other US key is absent locally and is ineligible for a Redis reread during the ten-second interval enforced between batch reads by \texttt{DualCache}, LiteLLM's local and Redis cache layer. Three runs contain a primary response followed later by fallback. The interval between Redis reads explains why stock alternates between primary and fallback responses.

\subsection{Identity propagation}\label{sec:identity-eval}
After a single attacker burst, four concurrent victim tenants experience $100\%$ denial with stock cooldown and $0\%$ with tenant scoping, with $0/120$ cross-thread identity mismatches. Under asynchronous execution, all $560$ cache reads and writes carry the executing task's registered tenant identity, and all six concurrent victim tenants are served. Tenant identity is stored in request context. A Redis variant also preserves attribution and produces tenant keys for $429$ writes.

A separate single-worker proxy test uses mocks, sequential requests, and no per-key quota. A custom authentication callback validates bearer keys against a server-side mapping. A distinct authenticated victim is denied $20/20$ on the recovered pool with stock cooldown and served $20/20$ with tenant scoping. All $111$ cache operations match the authenticated tenant associated with their request ID, and Redis contains cooldown state for the attacker's tenant alone.

For asynchronous requests, tenant attribution is checked against the identity registered for the executing task. For the proxy runs, request and process IDs link cache operations to access logs and authenticated identities (Table~\ref{tab:joint-repeat}). Sequential Router tests supply a trusted tenant identity directly.

\subsection{Additional configuration and cost measurements}\label{app:additional-cost}
\noindent\textbf{Repeated policy comparison.} We compare stock $12$\,s, short $1$\,s, and tenant-scoped $12$\,s cooldown in five independent repetitions. Each arm uses two mock deployments and a fresh process. A deployment tracks a two-second sliding window of all arriving requests, including $429$ rejections, and admits a request when fewer than four previous arrivals remain. A single thread launches requests against absolute deadlines $0.05$\,s apart. The four-tenant schedule sends $15$ consecutive requests per tenant in order. These Router experiments use a supplied tenant identity.

Raising the failure threshold changes the burst needed to trigger cooldown. A sweep with one run per setting sets the rate-limit failure threshold for each deployment to $12$. Its two primary mocks each use an eight-request, two-second sliding window, with a $15$\,s cooldown and retries disabled. With fallback disabled, the first successful victim samples occur at $3$\,s after a $20$-call burst and $17$\,s after an $80$-call burst.

Removing $429$ cooldown entirely allows more attempts against saturated capacity. We quantify this cost with overload, retry, and outage controls; every arrival, including rejected attempts, remains in the mock's sliding window. Skipping $429$ cooldown during a saturating $60$-call burst generates $6\times$ stock's upstream requests ($+50$ upstream $429$s), rising to $16\times$ with two retries per request. Both policies serve the $8$ requests admitted by the window, so extra attempts yield rate-limit errors. The service benefit appears after recovery, when stock still excludes the backend. Retaining fault back-off for a genuine $5xx$ outage yields $1.0\times$ stock upstream load.

Under a separate schedule with continuous saturation, scoped denial is $61\%$ versus $67\%$ for stock. The comparisons with multiple tenants and after recovery under the same schedules appear in \S\ref{sec:baseline}.

\subsection{Concurrent proxy load, latency, and Redis memory}\label{sec:overhead}\label{sec:concurrent-proxy}
We evaluate tenant scoping in five paired repetitions against stock LiteLLM v1.99.0. Each arm starts fresh PostgreSQL, Redis, and a four-worker authenticated proxy on one Apple M4 macOS host with $16$\,GiB memory. The configuration has $14$ deployments: a healthy pair and separate pairs for each fault condition. It uses simple-shuffle routing, $30$\,s cooldown, no retries or fallback, and $10{,}000$\,rpm per database key. Arm order alternates across repetitions; condition order is seeded and matched. All measured healthy requests succeed, with requests reaching all four workers in every run.

\noindent\textbf{Concurrent fault discovery.} For each of $4$, $16$, and $64$ tenants, four waves issue one request per tenant, with at most $32$ requests in flight. The selected mock pair continuously returns $429$ or $500$. The mock then switches to $200$, and a previously idle key sends eight concurrent recovery probes. Each error/count condition has a separate deployment pair. Table~\ref{tab:concurrent-fault} reports the five repetitions. With $64$ tenants and $429$ failures, scoping raises median upstream requests from $15$ to $209$, while restoring all $40$ victim probes after recovery.

The $500$ workloads retain global cooldown and show different recovery outcomes. At $4$ tenants, both arms make $16$ upstream requests in every run. Successful recovery probes are $8,8,8,8,3$ for stock and $2,3,2,3,3$ for scoping, out of eight per run. Thus, scoping serves fewer probes in four pairs and ties in one, for totals of $35/40$ and $13/40$. The cause of this difference remains unresolved. At $64$ tenants, both arms deny all $40$ immediate probes. These observations measure access immediately after the backend switches to $200$; fault waves use no retries.

\begin{table}[htbp]\centering\small
\caption{Concurrent fault workloads on the authenticated proxy. Upstream requests count HTTP POSTs reaching the backend, reported as medians [observed min--max] across five runs. Each run issues four requests per tenant and eight victim probes after recovery.}\label{tab:concurrent-fault}
\begin{tabularx}{\linewidth}{@{}ccLLLL@{}}\toprule
Error & Tenants & Stock requests & Scoped requests & Stock victim served & Scoped victim served \\\midrule
429 & 4 & 8 [6--8] & 11 [10--12] & 0/40 & 40/40 \\
429 & 16 & 14 [12--18] & 45 [39--56] & 0/40 & 40/40 \\
429 & 64 & 15 [13--20] & 209 [171--228] & 0/40 & 40/40 \\
500 & 4 & 16 [16--16] & 16 [16--16] & 35/40 & 13/40 \\
500 & 16 & 34 [30--38] & 27 [20--35] & 3/40 & 4/40 \\
500 & 64 & 42 [34--43] & 44 [43--50] & 0/40 & 0/40 \\\bottomrule
\end{tabularx}
\end{table}

\noindent\textbf{Latency.} Normal traffic uses tenant/concurrency pairs of $1/1$, $16/16$, and $64/32$. Each condition warms up with at least $64$ requests, then measures $1024$ requests with shuffled tenant order. Latency starts when a request acquires a client concurrency slot and ends at the HTTP response. Table~\ref{tab:concurrent-latency} reports medians of the five run-level percentiles; all $30{,}720$ measured requests return $200$. At $16$ tenants, the median p50 increases from $19.35$ to $25.10$\,ms. At $64$ tenants, run-level p99 spans $278.55$--$471.88$\,ms for stock and $405.84$--$475.08$\,ms for scoping. These measurements include local-cache behavior: most healthy requests avoid a network cache lookup. Forced cache misses and production traffic are outside this measurement.

\begin{table}[htbp]\centering\small
\caption{Healthy proxy latency. Values are medians across five runs, each measuring $1024$ requests per policy and workload.}\label{tab:concurrent-latency}
\begin{tabular*}{\linewidth}{@{\extracolsep{\fill}}rrrrrr@{}}\toprule
\multicolumn{2}{c}{\textbf{Workload}} & \multicolumn{2}{c}{\textbf{Median (ms)}} & \multicolumn{2}{c}{\textbf{99th percentile (ms)}} \\
\cmidrule(lr){1-2}\cmidrule(lr){3-4}\cmidrule(l){5-6}
Tenants & \shortstack{Concurrent\\requests} & Stock & \shortstack{Tenant\\scoped} & Stock & \shortstack{Tenant\\scoped} \\\midrule
1 & 1 & 3.23 & 3.25 & 5.48 & 14.68 \\
16 & 16 & 19.35 & 25.10 & 52.56 & 54.72 \\
64 & 32 & 80.24 & 81.74 & 430.57 & 418.66 \\\bottomrule
\end{tabular*}
\end{table}

\noindent\textbf{Redis memory.} A separate population measurement writes $429$ records for two deployments and $4$, $16$, $64$, $256$, and $1024$ tenants, using $300$\,s TTLs. Memory is the sum of Redis entry allocations after the writes. Scoped counts are $8$, $32$, $128$, $512$, and $2048$ entries, respectively. At $1024$ tenants, these entries occupy $376{,}832$ bytes ($368$\,KiB); stock retains two entries occupying $368$ bytes. These are measured entry allocations on this Redis instance, excluding process and client overhead. The counts follow the tenant/deployment product in \S\ref{sec:defense}.

\section{Configuration Analysis}\label{sec:general}
Configuration determines both which requests share exclusion state and which failures can update it. The AWS sample illustrates how deployment IDs determine this sharing, while Kong and Portkey use different routing controls.

\subsection{Fallback independence at the cache boundary}
Let $S$ and $F$ be the sets of primary and fallback entries in the model list that pass the victim's filters other than cooldown. An entry $e\in S\cup F$ belongs to either set, and $\kappa(e)$ is its actual cooldown key. For any set of such entries $X\subseteq S\cup F$, define $K(X)=\{\kappa(e):e\in X\}$ as the set of their distinct cooldown keys. Thus $K(S)$ and $K(F)$ are the primary and fallback key sets. Suppose all keys in $K(S)$ are active. With $f$ denoting a fallback entry, define
\[
 F_{\mathrm{outside}}=\{f\in F:\kappa(f)\notin K(S)\}.
\]
The set $F_{\mathrm{outside}}$ contains fallback entries whose keys are absent from $K(S)$, as indicated by $\notin$; the subscript describes this condition. If $K(F)\subseteq K(S)$, every fallback key is also a primary key. Cooling the primary keys then excludes the entire fallback set, even when its aliases or configured quotas differ. A key outside $K(S)$ provides a structurally independent candidate; service additionally requires that candidate to be uncooled, compliant, and physically available. Distinct cooldown keys therefore determine independence from the primary's exclusion state.

\subsection{A public gateway configuration}
AWS's resilience example combines pool routing, fallback, and consumer quotas~\citep{awsresilience,awsresilienceconfig}. It uses LiteLLM~1.89.1~\citep{litellm1891} with a single proxy worker. In the example, deployment entries retain explicit IDs or receive IDs derived from the model group name and ordered provider parameters. These IDs determine which entries share cooldown state.

\noindent\textbf{Shared primary and fallback identity.} The sample has eight entries in the model list and seven distinct cooldown IDs. A $3$\,rpm Haiku entry in the primary group of two entries and its $25$\,rpm fallback reuse the same explicit deployment ID. Both consult the same \texttt{deployment:<id>:cooldown} record. Cooling that ID excludes both entries despite their different configured limits. The sample's other fallback edge has different generated IDs and adds an independent cooldown key. Table~\ref{tab:config-families} records both outcomes.

\noindent\textbf{The trigger depends on the rejection stage.} The sample enables \texttt{simple-shuffle} and pre-call checks. Its local RPM filter runs before deployment selection; when every candidate exceeds its limit, it raises \texttt{RouterRateLimitErrorBasic} and exits before the deployment cooldown writer. A selected provider call returning $429$ carries deployment identity into the failure callback. The configured \texttt{allowed\_fails}$=2$ selects a path that counts failures: starting from an empty counter, a third eligible failure callback requests cooldown if the preceding count remains live. Retries can produce multiple callbacks per admitted client request. Triggering shared cooldown thus requires eligible failures attributed to a deployment. The database experiment uses usage-based routing and exercises a pre-call rejection attributed to a deployment; its error path carries that identity into cooldown accounting.

\noindent\textbf{Consumer identity and reachability.} The three consumer aliases receive distinct generated IDs, so traffic confined to each alias consults separate cooldown keys. The supplied configuration leaves authentication and access control from keys to aliases unspecified. With no externally supplied master key or alternative authentication, the sample accepts bearer strings without authenticating separate tenants. The proxy experiment in \S\ref{sec:quota} shows an authenticated tenant triggering the attack within its quota.

\begin{table}[t]\centering\small
\caption{Conditions governing shared exclusions. LiteLLM controls and Kong provide runtime observations; AWS rows give static key relationships and rejection paths. Full-pool denial additionally requires every compliant, serviceable alternative to be excluded.}\label{tab:config-families}
\begin{tabularx}{\textwidth}{@{}>{\RaggedRight\arraybackslash}p{3.15cm}LL@{}}\toprule
Configuration / evidence & Condition examined & Supported consequence and boundary \\\midrule
Rejected LiteLLM requests (\S\ref{sec:origin-attack}) & Attribution: registered deployment IDs versus absent or unknown IDs. & Two rejected requests with registered IDs reroute $B$, with no upstream calls from those requests. Negative controls leave $B$ on primary. IDs are supplied by the harness. \\
Shared group, distinct DB keys (\S\ref{sec:quota}) & $15$\,rpm/key; two $6$\,rpm deployments; rejection attributed to a deployment. & $A$'s $14$ admitted calls reroute $B$ in the measured configuration. \\
Tenant-scoped LiteLLM (\S\ref{sec:joint-eval}) & Sharing: $429$ records are indexed by deployment and authenticated tenant. & All $60$ victim requests reach recovered primaries; stock sends $54/60$ to fallback. Global error records remain shared. \\
AWS primary / fallback Haiku & $3$ and $25$\,rpm entries reuse one explicit ID. & Primary and fallback share one cooldown key; a live write excludes both. \\
AWS alternative fallback edge & Primary and fallback have different generated IDs. & One key outside the primary set; surviving other filters remains necessary. \\
AWS consumer aliases & Three distinct generated IDs; authorization from keys to aliases unspecified. & Traffic confined to each alias has a separate cooldown view; enforcing tenant confinement requires authentication and access control. \\
AWS local RPM rejection & Preselection filter raises an error for the group. & Rejection exits before the deployment cooldown writer. \\
Kong shared passive health & Lifetime: users share backend health records; passive failure threshold enabled. & Measured exclusion after recovery; active checks restore service after approximately $9$\,s. Passive checking is off by default. \\\bottomrule
\end{tabularx}
\end{table}

\subsection{Other gateways}
\noindent\textbf{Kong.} Kong~\citep{kong} stores passive health in the shared-memory dictionary \texttt{kong\_healthchecks}, keyed per target and consulted by the balancer. Our Kong~3.6 experiment is DB-less and containerized, with two authenticated consumers and a shared pool of two mock targets. Requests are sequential, with no per-key quota. Passive health checking is off by default (\texttt{http\_failures}$=0$). We enable a threshold of three failures; $429$ already belongs to the default list of unhealthy statuses for passive checks alongside $500$ and $503$~\citep{konghealth}.

Consumer $A$'s burst marks the shared targets unhealthy. A direct target probe returns $200$, while the distinct consumer $B$ receives HTTP~503 with no peer available in the ring-balancer (Table~\ref{tab:kong}). With active checks off, exclusion persists in the observed run; enabling active checks restores $B$ after approximately $9$\,s. Sharing target health state creates the dependency between consumers, while recovery configuration controls its persistence.

\noindent\textbf{Portkey.} Portkey~\citep{portkey} supports input and output guardrails and configurable fallback on guardrail failure~\citep{portkeyguardrails}. These checks operate on request and response content. Our dilemma depends on the selected destination's compliance with the tenant's deployment policy, such as residency or model identity. Portkey's enforcement of destination policies remains unverified in this study.

\begin{table}[t]\centering\small
\caption{Denial affecting another consumer in Kong 3.6 with passive health checking enabled. A direct probe serves while the gateway excludes the healthy target.}\label{tab:kong}
\begin{tabularx}{\columnwidth}{@{}LL@{}}\toprule
Stage & Result \\\midrule
consumer $B$ before attack & served ($200$) \\
consumer $A$ sends $40$ requests & $429$s trigger passive health exclusion \\
direct probe to target (bypass Kong) & healthy ($200$) \\
consumer $B$ after $A$'s burst & \textbf{[\texttt{503}] denied} (no peer in ring-balancer) \\
consumer $B$ after ${\sim}9$\,s (active checks on) & service restored ($200$) \\\bottomrule
\end{tabularx}
\end{table}

\end{document}